\documentstyle[12pt]{article}
\if@twoside \oddsidemargin 21pt \evensidemargin 59pt
\else \oddsidemargin 5pt \evensidemargin 5pt
\fi
\advance\textheight by \topskip
\newcommand {\beq}{\begin{equation}}
\newcommand {\eeq}{\end{equation}}
\newcommand {\beqa}{\begin{eqnarray}}
\newcommand {\eeqa}{\end{eqnarray}}
\renewcommand {\d}{d\!\!\!/}
\newcommand {\dd}{d\!\!\!\!/\!\!/}

\newcommand {\eq}[1]{(\ref{#1})}
\newcommand {\n}{\nonumber \\}
\newcommand {\nab}{\nabla\hspace*{-4pt}{}_{\mu}}
\newcommand {\cleqn}{\setcounter{equation}{0}}

\begin  {document}
\title {
{\normalsize\hfill\parbox{30mm}{
   KUNS~1248\\
   HE(TH)94/02 \\
   February, 1994\\
   hep-th/9403071}
   }
{}~~\\~\\~\\
         $N=2~$ Supersymmetry in Exterior Form
   }
\author {~\\~\\
         Masato Shimono\thanks{
          E-mail address: shimono@jpnyitp.bitnet.}  ~\\~\\}
\date   {{\it Department of Physics,
         Kyoto University \\
         Kitashirakawa, Sakyo-ku, Kyoto 606, Japan} ~\\~\\~\\}
\maketitle
\begin{abstract}
\indent
    An $N=2$ supersymmetric model using K\"{a}hler fields
    is proposed.
    It is a modified version of two-dimensional Benn-Tucker model.
    It indicates a geometrical origin
    of $N=2$ supersymmetry.
\end{abstract}
\newpage
\cleqn
\indent
K\"{a}hler proposed~\cite{k,g}
a description of fields with half-integral spin
in terms of differential forms.
All physical fields are described by
inhomogeneous differential forms in his framework
and Clifford product between differential forms
is introduced for economical description.
Especially (massless) fermionic K\"{a}hler field $\psi$ obeys
\beq
  \d\psi=dx^{\mu}\nab\psi=(d+d^*)\psi=0~,
 \label{1}
\eeq
which is called K\"{a}hler-Dirac equation (KDE, in short).
Here $\nab$ is the covariant derivative,
$d$ is the exterior differential, $d^*=*^{-1}d*$ is its adjoint
and juxtaposition denotes the Clifford multiplication.
KDE generically describes $2^{[\frac{n}{2}]}$-fold
degenerate fermions in $n$-dimensional space-time.
The degeneracy can be removed
by ideal decomposition~\cite{g,bt3} of K\"{a}hler field,
which does not work in general curved space-time,
whilst there are attempts to interpret it
as an origin of fermion \lq generation'~\cite{bdh,bt2}.

KDE has been studied from miscellaneous aspects, that is,
supersymmetric extensions~\cite{gurs}--\cite{sc},
lattice theories~\cite{sc}--\cite{shimo},
Kaluza-Klein theories~\cite{bt1}--\cite{bul2},
phenomenological considerations~\cite{h,kr},
quantization procedures~\cite{bt3,bht,j}
and possible modifications for
curved manifolds~\cite{bt3,bt2,h,bul2}.
Among them, Benn and Tucker proposed~\cite{bt4}
an attractive model which has a certain Fermi-Bose symmetry
described by K\"{a}hler fields.
Hereafter we shall follow the conventions of refs.\cite{bt4}.

Let
$\psi
$
and
$\phi
$
be real Grassmann odd and even valuables, respectively,
(even though the statistics are not necessarily determined, )
the Benn-Tucker model is described by the action 
\beqa
    S&=&\int \!\!z~ S_0\left(
            \bar{\psi}  \d \psi   B
           +\bar{\phi}  \d(\d+\dd)\phi
                  \right)~.
   \label{2}
\eeqa
$S_p(\psi)$ means the $p$-form part of $\psi$,
$z=\sqrt{g}dx^1\cdots dx^{n}$ is the volume $n$-form and
$B$ is some covariantly constant form which satisfies
$\bar{B}\equiv \zeta\eta B=-B$.
Here automorphism $\eta$ and antiautomorphism $\zeta$ are defined by
\beqa
  \eta S_p(\phi)=(-1)^{p}S_p(\phi)~,
 \label{3} \\
  \zeta S_p(\phi)=(-1)^{[\frac{p}{2}]}S_p(\phi)~.
 \label{4}
\eeqa
We have also defined as
$\dd = -\zeta\d\zeta$,
which satisfies $\overline{\d \psi}=\dd \bar{\psi}$.
In two dimensions, for example, this action describes
a scalar particle, a pseudoscalar particle,
a vector particle and two spin-$\frac{1}{2}$ particles.

In any dimensions this action \eq{2} is invariant
under the following supersymmetry-like transformations:
\beqa
   Q_{\alpha} \psi &=& (\d+\dd)\phi_{\pm} \alpha~,
  \label{5}
   \\
   Q_{\alpha} \phi_{\pm} &=& \left(\psi B \bar{\alpha}\right)_{\pm}~,
  \label{6}
\eeqa
where the parameter $\alpha$ is
a Grassmann odd covariantly constant form,
and $\phi_{\pm}=(1\pm\eta)\phi$ is the even (odd) form part.
This action is also invariant under gauge transformations
\beqa
  \phi_+ &\rightarrow& \phi_+ + d^*\alpha_-~,
 \label{7} \\
  \phi_- &\rightarrow& \phi_- + d\alpha_+~,
 \label{8}
\eeqa
for some even (odd) covariantly constant form $\alpha_{\pm}$.

In four dimensions,
if we suppose that the fermionic forms $\psi, \alpha$ and $\beta$
are all in a minimal left ideal of some projection operator $P$
which satisfies $\bar{P}B=BP$,
i.e., $\psi P=P$ and so on,
the commutator of these transformations
produces~\cite{bt4} a parallel transport
modulo gauge transformations \eq{7} and \eq{8}:
\beq
  {}~[Q_{\alpha},Q_{\beta}]=L_K ~.
 \label{9}
\eeq
Here $K$ denotes a Killing vector which is dual to
$-4S_1(\alpha B \bar{\beta})$
and $L_K$ denotes the Lie derivative along the direction of $K$.
That is, the transformations \eq{5} and \eq{6}
satisfy the ordinary $N=1$ superalgebra.

Although the off-shell closure \eq{9} of superalgebra
of four-dimensional Benn-Tucker model completely
depends on its dimensionality~\cite{bt4},
we can obtain similar superalgebra 
in its two-dimensional version
after a slight modification.
There we do not need to split the bosonic field $\phi$
into even and odd parts.
Then supersymmetric transformations take the following forms:
\beqa
   Q_{\alpha} \psi &=& (\d+\dd)\phi \alpha~,
  \label{10}
   \\
   Q_{\alpha} \phi &=& \psi B \bar{\alpha}~.
  \label{11}
\eeqa
If $\psi, \alpha$ and $\beta$ are all in a minimal left ideal,
the commutator also produces
a parallel transport modulo gauge transformations.
That is, it produces
\beq
  {}~[Q_{\alpha},Q_{\beta}]\psi=
    -4S_0(\alpha B \bar{\beta} dx^{\mu})\nab\psi
 \label{12}
\eeq
for the fermionic field $\psi$, and
\beqa
  {}~[Q_{\alpha},Q_{\beta}]\phi&=&
    -4S_0(\alpha B \bar{\beta} dx^{\mu})\nab\phi \n
    &&\hspace{3mm}
    +4d\left[
        \phi\left(
         S_1(\alpha B \bar{\beta})+S_2(\alpha B \bar{\beta})
         \right)
       \right]_+
    +4d^*\left[
          \phi\left(
           S_1(\alpha B \bar{\beta})+S_2(\alpha B \bar{\beta})
          \right)
         \right]_- \n
 \label{13}
\eeqa
for the bosonic field $\phi$.

Since two-dimensional K\"{a}hler fields contain
two degenerated fermion species,
one may hope the model should possess $N=2$ supersymmetry
without restriction on fermionic variables
to a minimal left ideal.
Unfortunately it is not the case in a naive way,
for the off-shell closure of two-dimensional superalgebra
is due to the fact that the transformations \eq{10} and \eq{11}
contain the full bosonic degrees of freedom
in comparison with the four-dimensional ones
using only a half of bosonic degrees.
Therefore, in order to offer the sufficient bosonic degrees of freedom
for $N=2$ supersymmetry,
we must introduce a bosonic field $\varphi$,
and add a new term
to the action \eq{2}.
Now the action
\beqa
    S&=&\int \!\!z~ S_0\left(
            \bar{\psi}  \d \psi   B
           +\bar{\phi}  \d(\d+\dd)\phi
           +\bar{\varphi}  \d(\d-\dd)\varphi
          \right) ~
  \label{15}
\eeqa
has additional field equations
\beqa
  d^*d\varphi_+ &=& 0 ~,
 \label{16} \\
  dd^*\varphi_- &=& 0 ~,
 \label{17}
\eeqa
which yield a new gauge invariance under transformations
\beqa
  \varphi_+ &\rightarrow & \varphi_+ + d\alpha_-~,
 \label{18} \\
  \varphi_- &\rightarrow & \varphi_- + d^*\alpha_+~.
 \label{19}
\eeqa
Then we can find a new type of supersymmetric transformations
which leaves the action invariant:
\beqa
   Q_{\alpha} \psi
    &=& (\d+\dd) \phi \alpha
        +(\d-\dd) \varphi \alpha B~,
  \label{20} \\
   Q_{\alpha} \phi &=& \psi B \bar{\alpha}~,
  \label{21} \\
   Q_{\alpha} \varphi &=& -\psi B^2 \bar{\alpha}~.
  \label{22}
\eeqa
In this case, if we suppose $B^2=1$, we have $N=2$ superalgebra:
\beqa
  {}~[Q_{\alpha},Q_{\beta}]\psi
    &=&-4S_0(\alpha B \bar{\beta} dx^{\mu})\nab\psi \n
    &=&-4\sum_{i=1,2}S_0
        (\alpha_i B \bar{\beta}_i dx^{\mu})\nab\psi ~,
 \label{23} \\
  {}~[Q_{\alpha},Q_{\beta}]\phi&=&
    -4\sum_{i=1,2}S_0
        (\alpha_i B \bar{\beta}_i dx^{\mu})\nab\phi \n
    &&\hspace{3mm}
    +4d\left[
        \varphi S_0(\alpha \bar{\beta})
        +\phi\left(
         S_1(\alpha B \bar{\beta})+S_2(\alpha B \bar{\beta})
         \right)
       \right]_+ \n
    &&\hspace{3mm}
    +4d^*\left[
          \varphi S_0(\alpha \bar{\beta})
          +\phi\left(
           S_1(\alpha B \bar{\beta})+S_2(\alpha B \bar{\beta})
          \right)
         \right]_-~,
 \label{24} \\
  {}~[Q_{\alpha},Q_{\beta}]\varphi&=&
    -4\sum_{i=1,2}S_0
        (\alpha_i B \bar{\beta}_i dx^{\mu})\nab\varphi \n
    &&\hspace{3mm}
    -4d\left[
        \phi S_0(\alpha \bar{\beta})
        +\varphi\left(
         S_1(\alpha B \bar{\beta})+S_2(\alpha B \bar{\beta})
        \right)
       \right]_- \n
    &&\hspace{3mm}
    -4d^*\left[
          \phi S_0(\alpha \bar{\beta})
           +\varphi\left(
           S_1(\alpha B \bar{\beta})+S_2(\alpha B \bar{\beta})
          \right)
         \right]_+~.
 \label{25}
\eeqa
Here $\alpha_i=\alpha P_i$ and $\beta_i=\beta P_i$ are
minimal left ideals for some projection operators $P_1$
and $P_2=1-P_1$ which satisfy $\bar{P}_iB=BP_i$.
Namely, as well as the fermion field $\psi$,
all fermionic parameters generically have two-fold degeneracy.
That is a possible origin of $N=2$ supersymmetry,
as has already been suggested by Scott~\cite{sc}.

Here we have proposed a model with $N=2$ supersymmetry,
which might have its origin
in the inherent multiplicity of fermion species of K\"{a}hler fields.
Similar investigations in higher dimensions
together with
Kaluza-Klein context for
K\"{a}hler fields~\cite{bul2}
are attractive tasks.
Application of this model to lattice theories
is also intriguing and will be discussed elsewhere~\cite{shimo2}.
Another interesting possibility is an attempt~\cite{bul3} to extend
global supersymmetry of models with K\"{a}hler fields
into {\it local} one.
This problem is closely related with
the peculiar nature of K\"{a}hler fields
in curved space-time~\cite{bdh,bt2,bul2}
and (non-)existence of a covariantly constant Killing vector
in general curved space-time~\cite{bt4,bpt}.
The naive KDE should be modified appropriately there.
However, we hope that the investigation along this line
should bring us a new insight into
a relation between Fermi-Bose symmetry
and geometrical property of space-time manifold.

\section*{Acknowledgements}
\indent\indent
I am very grateful to T.~Maskawa
for his valuable comments
and for his persistent encouragement.
I also wish to thank S.~Yahikozawa for instructive discussions.

{}~~

{}~~

\noindent
{\it Note added.} \indent
After completing this work, I was informed of
a series of papers \cite{agz}, in which similar,
but different supersymmetric models are proposed
both in continuum and lattice theories.

\end{document}